\documentstyle[11pt,epsfig]{article}
\newcommand{\beq}{\begin{equation}}
\newcommand{\eeq}{\end{equation}}
\newcommand{\bea}{\begin{eqnarray}}
\newcommand{\eea}{\end{eqnarray}}
\def\lsim{\raise0.3ex\hbox{$\;<$\kern-0.75em\raise-1.1ex\hbox{$\sim\;$}}}
\def\gsim{\raise0.3ex\hbox{$\;>$\kern-0.75em\raise-1.1ex\hbox{$\sim\;$}}}
 
\begin{document}
\begin{center}
{\large{\bf Neutrinos from Decaying Muons, Pions, Neutrons and Kaons in Gamma Ray Bursts}}\\ 
\medskip
{Reetanjali Moharana \footnote{Email: reetanjali@iitb.ac.in,}, Nayantara Gupta \footnote{Email: nayan@rri.res.in}}\\
{$^1$Department of Physics, Indian Institute of Technology Bombay, Mumbai 400 076, India\\
$^2$Astronomy \& Astrophysics Group, Raman Research Institute, C.V. Raman Avenue, Bangalore, India}
\end{center}

\begin{abstract}
In the internal shock model of gamma ray bursts ultrahigh energy muons, pions,
neutrons and kaons are likely to be produced in the interactions of shock accelerated relativistic protons with low energy photons (KeV-MeV). These particles 
subsequently decay to high energy neutrinos/antineutrinos and other secondaries.
 In the high internal magnetic fields of gamma ray bursts, the ultrahigh energy charged particles ($\mu^+$, $\pi^+$, $K^+$) lose energy significantly due to synchrotron radiations before decaying into secondary high energy neutrinos and antineutrinos. The relativistic neutrons decay to high energy antineutrinos, protons and electrons. We have calculated the total neutrino flux (neutrino and antineutrino) considering the decay channels of ultrahigh energy muons, pions, neutrons and kaons. We have shown that the total neutrino flux generated in neutron decay can be higher than that produced in $\mu^+$ and $\pi^+$ decay. The charged kaons being heavier than pions, lose energy slowly and their secondary total neutrino flux is more than that from muons and pions at very high energy. Our detailed calculations on secondary particle production in $p\gamma$ interactions give the total neutrino fluxes and their flavour ratios expected on earth. Depending on the values of the parameters (luminosity, Lorentz factor, variability time, spectral indices and break energy in the photon spectrum) of a gamma ray burst the contributions to the total neutrino flux from the decay of different particles (muon, pion, neutron and kaon) may vary and they would also be reflected on the neutrino flavour ratios. 
\end{abstract}

\section{Introduction}
Extensive airshower arrays \cite{pa1,pa2,hires,agasa,yakutsk} have detected a large number of ultrahigh energy cosmic ray events.
The origin of these cosmic rays are yet to be identified.
They may come from supernova remnants (SNRs), active galactic nuclei (AGN), gamma ray bursts (GRBs) or some unknown sources. 
Some of the ultrahigh energy cosmic rays (protons and nuclei) may interact with low energy radiations and matter inside the source producing secondary 
high energy photons and neutrinos. Protons can be shock accelerated to $\sim 10^{21}$ eV inside GRBs by Fermi mechanism. 
Shock accelerated protons may loose energy by synchrotron cooling and $p\gamma$ interaction inside the source depending on the magnetic field and 
low energy photon density respectively. The high energy neutrino flux produced in $p\gamma$ interactions have been calculated in detail in many earlier papers \cite{vietri,wax1,rach,alv,nay1,guetta,bhat,nay2,murase1,becker1,wax2,murase2}. 
It has been noted earlier that $pp$ interactions are only important for photospheric radii of GRB fireballs \cite{murase3,dai}.
The interactions of shock accelerated protons with low energy radiations ($p\gamma$) lead to the production of ultrahigh energy mesons 
and leptons
 $p\gamma\rightarrow{\pi^{+,0}}X,\, {\pi^{+}} \, \rightarrow \,
 {\mu^{+}}\nu_{\mu}, \, {\mu^{+}} \, \rightarrow \, {e^{+}}
 \bar\nu_{\mu}\nu_e$,
$X$ can be neutrons and they will decay to protons, electrons and antineutrinos of electron flavour ($n\,\rightarrow p +e^-+\overline{\nu}_e$) in 886 sec in their rest frames. 
The other secondary products in $p\gamma$ interactions are 
$p\gamma \rightarrow{K^{+,0}}X$ where X can be either $\Lambda^0$, $\Sigma^0$ or $\Sigma^+$. 
Kaons decay to lighter mesons, leptons and neutrinos.   
If the magnetic field inside the GRBs is very high then the ultrahigh energy muons, pions and kaons lose energy significantly before decaying 
to neutrinos. Synchrotron cooling of charged muons and mesons would lead to decrease in the fluxes of neutrinos from their decay. 
Although, the cross-section for kaon production from p$\gamma$ is much less than that for photo-pion production, kaons being heavier cool at a much slower rate 
compared to muons and pions. As a result at very high energy the kaon decay channel of neutrino production becomes significant compared to the pion decay channel \cite{and,asan,wal1,wal2,kal1}. 
Shock accelerated protons are expected to produce high energy neutrons in various interactions, and these unstable particles
 subsequently decay to leptons and protons. 

In the context of UHECR production from GRBs the neutron decay channel was considered earlier in \cite{der}. The energy and time dependence of neutron, neutrino fluxes from GRBs were calculated within the scenario of external shock model in this paper. Also the radiation halos from neutron decay electrons, protons and their detectability in different wavelengths were studied. 

In a recent paper \cite{ahl} the authors have explored the possiblity of detecting high energy diffuse neutrino flux from GRBs by IceCube, assuming the diffuse UHECR flux as detected by HiRes is the proton flux produced by the decay of cosmic ray neutrons in GRB fireballs.

In our work we have not assumed that GRBs are the sources of the  
UHECR events observed by Pierre Auger \cite{pa1,pa2}, HiRes \cite{hires} or 
AGASA \cite{agasa} experiments. We have considered the possibility of cosmic ray acceleration in individual GRBs to ultra high energies to explore the parameter dependence of the high energy neutrino fluxes produced in decay of various particles. We discussed the importance of the neutron decay channel in our earlier 
paper \cite{ree} for cosmic accelerators with high internal magnetic fields.
In this paper we have calculated the neutrino fluxes from GRBs in the internal shock model through muon, photo-pion, neutron and kaon decay, 
including the effect of proton energy loss by synchrotron cooling and $p\gamma$ interactions, synchrotron cooling of charged muons, pions, 
and kaons in the internal magnetic field of GRB. 
After including proton energy losses, we find that the neutron decay channel of 
high energy antineutrino production may remain important in GRBs depending on the values of the GRB parameters. Moreover, the various decay processes may show distinct features in the neutrino flavour ratios.

 Due to the ongoing experimental activities to detect high energy neutrinos from GRBs \cite{abbasi0,abbasi1,abbasi2,abbasi3,abbasi4} this field has remained exciting. IceCube collaboration has claimed to reach the sensitivity of detecting neutrino flux from GRBs at TeV energy. IceCube opearted in a 40-string configuration from April 5, 2008 to May 20, 2009 \cite{abbasi1}. They considered 117 bursts and no events were detected above the atmospheric background. For cosmic neutrinos with an $E^{-2}$ energy spectrum an integral flux limit of $E^2\phi\leq 3.6\times 10^{-8} GeV cm^{-2} sec^{-1}sr^{-1}$ has been found in the energy interval of $2\times10^6-6.3\times10^{9} GeV$ \cite{abbasi2}. In particular the non-detection of neutrinos from GRBs \cite{abbasi4} has placed a tighter upper bound which is 3.7 times below the theoretical predictions \cite{wax1,rach, guetta, ahl} combining 40 and 59 strings. IceCube collaboration has concluded \cite{abbasi4} that GRBs are not the only sources of cosmic rays above energy 1 EeV or the efficiency of neutrino production is much lower than the current predictions. The values of the bulk Lorentz factor $\Gamma$ and the ratio of energies in protons to electrons have been constrained at the $90\%$ confidence level in their Fig.4.  
H\"ummer et al. \cite{hum} have recalculated the neutrino fluxes from GRBs and 
concluded that their result is significantly below the limit set by IceCube experiment operated in 40 strings.
It has been pointed out by Li \cite{Li} that the theoretical prediction of neutrino flux from GRBs in IceCube papers is an overestimation, and the non-detection of GRB neutrinos is consistent with the correct theoretical estimation of neutrino flux.
He et al. \cite{He} have calculated the neutrino flux from GRBs including proton energy losses and they have also suggested that the neutrino flux predicted theoretically in the papers by IceCube collaborations is an overestimation. In calculating the photon number density IceCube collaboration approximated the energy of all photons by the break energy of the photon spectrum. The recalculated
neutrino flux by He et al. \cite{He} from 215 GRBs observed by IceCube 40 and 59 string configurations is about $36\%$ of the $90\%$ confidence upper limit obtained by IceCube collaboration and it is consistent with non-detection of GRB neutrinos by IceCube detector. 
The objective of the present work is to show that the relative importance of the different channels (pion, muon, neutron and kaon decay) of neutrino production in GRBs depends on the values of the GRB parameters. We have compared our calculated flavour ratios with earlier work \cite{wax3,wal2}. After including muon energy loss and kaon decay channels in detail at high energy, our flavour ratios are lower than the constant value of $\phi_{\nu_{\mu}}/(\phi_{\nu_{e}}+
\phi_{\nu_{\tau}})=0.64$ found in earlier work.
\section{High energy neutrinos from gamma ray bursts}
Frames of references have been assigned as ``c'' for comoving or wind rest frame, ``p'' for proton rest frame. Quantities measured in the source rest frame are written without any subscript.
In this paper we consider the muon, photo-pion, neutron and kaon decay channels of high energy neutrino production in the internal shock model of GRB prompt emission. We have used the low energy photon flux typically observed by $Swift$ in the energy range of 1 KeV to 10 MeV to calculate the neutrino flux from individual GRBs.
The photon energy spectrum from a GRB can be expressed as a broken
power law with break at
 $\epsilon_{\gamma}^b$ in the source rest frame related to the break energy
in the comoving frame
 $\epsilon_{\gamma,c}^b$ as $\epsilon_{\gamma}^b=\Gamma\epsilon_{\gamma,c}^b$.
\begin{equation}
\frac{dn_{\gamma}}{d\epsilon_{\gamma,c}}
=A \left\{ \begin{array}{l@{\quad \quad}l}
{\epsilon_{\gamma,c}}^{-\gamma_1} &
\epsilon_{\gamma,c}<\epsilon_{\gamma,c}^b\\{\epsilon_{\gamma,c}^b}^{\gamma_2-\gamma_1}
{\epsilon_{\gamma,c}}^{-\gamma_2} & \epsilon_{\gamma,c}>\epsilon_{\gamma,c}^b
\end{array}\right.
\label{photon}
\end{equation}
$\gamma_1 < 2$, and $\gamma_2 > 2$. The normalization constant $A$
is related to the internal energy density $U$ by,
\[
A=\frac{U{\epsilon_{\gamma,c}^b}^{\gamma_1-2}}{[\frac{1}{\gamma_2-2}-\frac{1}
{\gamma_1-2}]}
\label{A_g}
\]

These photons are interacting with shock accelerated protons to
produce charged and neutral pions and kaons. The maximum energy of the shock accelerated protons in the GRB fireball 
can be calculated by comparing the minimum of the $p\gamma$ interaction time scale ($t_{p\gamma}$), p-synchrotron cooling time scale ($t_{syn}$) and dynamical 
time scale ($t_{dyn}$) of a GRB with the acceleration time scale ($t_{acc}$) 
of the protons as discussed in \cite{nay3}.
\begin{equation}
t_{acc}=min(t_{p\gamma},\,t_{dyn},\, t_{syn})
\end{equation}
The inverse of the time scale of $p\gamma$ interactions with protons of
energy $\epsilon_{p,c}$ in the comoving/wind rest frame leading to the production of secondary particles `a' is,

\begin{eqnarray}
t_a^{-1}(\epsilon_{p,c}) &\equiv& -\frac{1}{\epsilon_{p,c}}\frac{d\epsilon_{p,c}}{dt}\\
&=& \frac{c}{2\Gamma_{p,c}^2}\int\limits_{\epsilon_{\gamma, th}}^\infty d\epsilon_{\gamma,p}
\sigma_a(\epsilon_{\gamma, p}) \xi_a (\epsilon_{\gamma,p}) \, \epsilon_{\gamma,p}
\, \int\limits_{\epsilon_{\gamma,p}/2\Gamma_p}^\infty d\epsilon_{\gamma,c} \, \epsilon_{\gamma,c}^{-2} \, \frac{dn(\epsilon_{\gamma,c})}{d\epsilon_{\gamma,c}}
\end{eqnarray}
 where $\Gamma_{p,c}=\epsilon_{p,c}/m_pc^2$, and $\epsilon_{\gamma,th}$ is the threshold energy of photon in proton rest frame.
$\epsilon_{\gamma,p}$ and $\epsilon_{\gamma,c}$ are the photon energies in the proton rest frame and the comoving frame respectively. $\xi_a(\epsilon_{\gamma,p})$ is the average fraction of energy lost to the secondary 
particle by a proton of energy $\epsilon_{p,c}$. The expression for the low energy photon flux can be substituted from equ.(\ref{photon}).
The second integration gives,
\begin{equation}
 I_2=A
 \left\{ \begin{array}{l@{\quad \quad}l}
 \frac{1}{\gamma_1+1} \, \frac{\epsilon_{\gamma,p}}{2\Gamma_{p,c}}^{-\gamma_1-1} \, - \, {\epsilon_{\gamma,c}^b}^{-\gamma_1-1} \, \left[\frac{1}{\gamma_1+1}- \frac{1}{\gamma_2+1}\right] 
  &
 \frac{\epsilon_{\gamma,p}}{2\Gamma_{p,c}} <\epsilon_{\gamma,c}^b\\

\frac{1}{\gamma_2+1} \, \frac{\epsilon_{\gamma,p}}{2\Gamma_{p,c}}^{-\gamma_2-1} \, {\epsilon_{\gamma,c}^b}^{\gamma_2-\gamma_1}
 &
 \frac{\epsilon_{\gamma,p}}{2\Gamma_{p,c}} >\epsilon_{\gamma,c}^b
 \end{array}\right.
 \label{na1}
 \end{equation}
 The inverse of the time scale of $p\gamma$ interactions leading to the production of secondary particles `a' can be expresses as ,

\begin{equation}
t_a^{-1}(\epsilon_{p,c})=\frac{c}{2\Gamma_{p,c}^2}\int_{\epsilon_{\gamma, th}}^\infty d\epsilon_{\gamma,p}
 \sigma(\epsilon_{\gamma, p}) \xi(\epsilon_{\gamma,p}) \, \epsilon_{\gamma,p} \, I_2.
\label{I1}
\end{equation}
The fractional energy transferred from proton to the secondary particle 
in an internal shock of radius $r_d$ is $f_a=r_d/(\Gamma c t_a)$.
\subsection{Neutrinos from photo-pion decay}
 We have considered production of pions in $p\gamma$ interactions through the decay of resonant particle $\Delta^+$.
At the delta resonance both $\pi^{0}$ and $\pi^{+}$ have been assumed to be produced with equal probabilities. $\pi^{+}$ gets on the average
$20\%$ of the proton's energy.
The charged pions decay to muons and neutrinos. Finally the muons decay to electrons and neutrinos, antineutrinos. Each pion decay followed by muon decay gives two neutrinos, one antineutrino and one positron.
If the final state leptons share the pion energy equally then each neutrino carries $5\%$ of the initial proton's energy.
The minimum energy of the protons interacting with photons of energy 
$\epsilon_{\gamma}^b$ for the $\Delta$ resonance with cross section $\sigma_{\Delta} \sim 5 \times 10^{-28}$ $cm^2$ in the source rest frame is,
\begin{equation}
\epsilon_{p,\Delta}^b=1.3\times10^{7}\Gamma_{300}^2(\epsilon_{\gamma,MeV}^b)^{-1}
{\rm GeV}.
\label{EPDbreak}
\end{equation}
The first break in the energy spectrum of neutrino,
$\epsilon_{\nu,\pi}^b$ is due to the break in the photon spectrum at $\epsilon_{\gamma}^b$.
\begin{equation}
\epsilon_{\nu,\pi}^b=6.5\times10^5 \Gamma^2_{300}
\left(\epsilon_{\gamma,MeV}^b\right)^{-1} {\rm GeV}.
\label{Enupibreak}
\end{equation}
The fireball Lorentz factor $\Gamma_{300}= \Gamma/{300}$, photon luminosity $L_{\gamma,51}=L_{\gamma}/(10^{51} ergs \, /sec)$,
variability time $t_{v,-3}=(t_v/10^{-3}sec)$ are the important parameters of a GRB. \\
The minimum Lorentz boost factor of a GRB wind for observability of photon
of energy $\epsilon_t$ 
(energy measured in wind rest frame) depends on $L_{\gamma}$ and $t_v$ as mentioned in \cite{wax}.
 \begin{equation}
\Gamma \ge 250 \left[L_{\gamma,51}\left(\frac{\epsilon_t}{100 \, MeV}\right)t_{v,-3}^{-1}\right]^{1/6}.
\label{Lorentz}
\end{equation}
The internal energy density $U$ relates to photon luminosity, $L_\gamma=4\pi {r_d}^2 \Gamma^2 c U$. $r_d=\Gamma^2ct_v$ is the internal shock radius.
The charged pions and muons lose energy due to synchrotron radiation in 
  magnetic fields inside the sources before decaying to neutrinos. 
 The expression for magnetic field in the comoving frame is,

\begin{equation}
B_c=5.04 \times 10^{5} \, \epsilon_{B,-1}^{1/2} \, L_{\gamma,51}^{1/2} \, \Gamma_{300}^{{-3}} \,  t_{v,-3}^{-1} \, G
\end{equation}
$\epsilon_B$ and $\epsilon_e$ are the energy fractions carried by the magnetic field and the electrons respectively. 
Pion cooling energy is ten times higher than muon cooling energy. We would be overestimating the total neutrino flux if we use the pion cooling energy to derive
the second break energy in the neutrino spectrum as the number of neutrinos produced from muon decay is more.
The muon cooling break energy $\epsilon_{\nu,\mu}^s$ can be expressed as a function of
GRB parameters.
\begin{equation}
\epsilon_{\nu,\mu}^{s}=2.56\times10^{6}\epsilon_e^{1/2}\epsilon_B^{-1/2}
L_{\gamma,51}^{-1/2}\Gamma_{300}^4t_{v,-3}{\rm GeV}.
\label{Enumusyn}
\end{equation}
The total energy to be emitted by neutrinos of energy $\epsilon_{\nu,\pi}$ from photo-pion decay (considering muon and pion decay neutrinos together) 
in the source rest frame of a GRB is \cite{nay2},
\begin{equation}
 \epsilon_{\nu,\pi}^2\frac{dN_{\nu}}{d\epsilon_{\nu,\pi}}\approx\frac{3f_{\pi}}
{8\kappa}\frac{(1-\epsilon_e-\epsilon_B)}{\epsilon_e}E_{\gamma}^{iso}
 \left\{\begin{array}{l} 1 \hspace{2.cm} \epsilon_{\nu,\pi}<\epsilon_{\nu,\mu}^{s}\\
 \left(\frac{\epsilon_{\nu,\pi}}{\epsilon_{\nu,\mu}^{s}}\right)^{-2} \hspace{1.cm} \epsilon_{\nu,\pi}>\epsilon_{\nu,\mu}^{s}
\end{array}
 \right.
\label{totnupi}
\end{equation}
where $E_{\gamma}^{iso}$ is the total isotropic energy of the emitted
gamma-ray photons in the energy range of 1keV to 10MeV, which is
available from observations. It is the product of $L_{\gamma}$ with the duration of the prompt emission from the GRB. 
 In deriving eqn.(\ref{totnupi}) it has been assumed that the shock accelerated electrons in the internal shock model are cooling fast. They are in the radiatively efficient regime and their energy goes to the emitted photons.
The fraction $\epsilon_p=1-\epsilon_e-\epsilon_B$ of shock energy is assumed to be all in relativistic power law protons. This is an extreme assumption that the shocks are $100\%$ efficient in converting bulk relativistic shock energy into relativistic power law electrons, protons and fields. However, it is known from simulations that in shocks many of the shocked protons remain in a thermal peak, and only some unspecified fraction of the total number of protons ends up in the power law. This reduces the power law proton (and neutrino) flux by a multiplicative proton injection fraction $\eta_p\leq1$. 

The maximum energy of neutrinos from pion decay is approximately $5\%$ of 
the maximum energy of protons ($\epsilon_{p,max}$). 
$f_{\pi}$ is the fractional energy of a proton going to pion inside the GRB fireball. It can be expressed in terms of the parameters of a GRB.
\begin{equation}
f_{\pi} (\epsilon_p)= f_0^{\pi} \left\{\begin{array}{l@{\quad\quad}l}
\frac{1.34^{\gamma_1-1}}{\gamma_1+1}\left(\frac{\epsilon_p}{\epsilon_{p,\Delta}^b}\right)^{\gamma_1-1}& {\epsilon_p>\epsilon_{p,\Delta}^b}\\
\frac{1.34^{\gamma_2-1}}{\gamma_2+1}\left(\frac{\epsilon_p}{\epsilon_{p,\Delta}^b}\right)^{\gamma_2-1}&
{\epsilon_p < \epsilon_{p,\Delta}^b}
\end{array}
\right.
\label{fpi}
\end{equation}
where $f_0^{\pi}=\xi_{\pi}\frac{4.5L_{\gamma,51}}{\Gamma_{300}^4 \,
t_{v,-3}{\epsilon_{\gamma,MeV}^b}}\frac{1}{\big
[\frac{1}{\gamma_2-2}-\frac{1}{\gamma_1-2}\big ]}$ and $\xi_\pi=0.2$. $\kappa$ is a normalization factor in eqn.(\ref{totnupi}). 
The relativistic electrons produce the photons by synchrotron radiation and inverse compton scattering of low energy photons.
 Four orders of magnitude in photon energy corresponds to two orders of magnitude in the energy of the radiating charged leptons.
 The ultra-relativistic electrons have a power law spectrum, spectral index is assumed to be $-2.5$. This corresponds to $\kappa=1.8$ 
assuming photon fluence is proportional to neutrino luminosity. 
\subsection{Neutrinos from neutron $\beta$-Decay}

Ultrahigh energy neutrons are also produced in $p\gamma$ interactions along with pions and kaons \cite{der,ahl}.
This channel was considered to estimate TeV neutrino flux from Cygnus OB2 located about 1.7 kpc away from us \cite{anc2}. We follow a similar formalism. 
The neutrons (with Lorentz factor $\Gamma_n$) decay ($n\,\rightarrow p +e^-+\overline{\nu}_e$) to $\bar\nu_e$ with 
a decay mean free path $c\Gamma_n \bar{\tau}_n \,=\,10(\epsilon_n/EeV)$ Kpc.
 $\overline{\tau}_n\,=\,886$ seconds is the lifetime of a neutron in its rest frame and $\epsilon_n$ is its energy in the source rest frame. The fraction of a proton's energy lost to neutron production in the process $p\gamma\rightarrow \pi^+n$ at the $\Delta-$resonance \cite{ree} is,
\begin{equation}
f_n (\epsilon_p)= f_0^n \left\{\begin{array}{l@{\quad\quad}l}
 \frac{1.34^{\gamma_1-1}}{\gamma_1+1}\left(\frac{\epsilon_p}{\epsilon_{p,\Delta}^b}\right)^{\gamma_1-1}& {\epsilon_p>\epsilon_{p,\Delta}^b}\\
 \frac{1.34^{\gamma_2-1}}{\gamma_2+1}\left(\frac{\epsilon_p}{\epsilon_{p,\Delta}^b}\right)^{\gamma_2-1}&
{\epsilon_{p}< \epsilon_{p,\Delta}^b}
\end{array}
\label{fn}
\right.
\end{equation}
where $f_0^n=\xi_n\frac{4.5L_{\gamma,51}}{\Gamma_{300}^4 \,
t_{v,-3}{\epsilon_{\gamma,MeV}^b}}\frac{1}{\big
[\frac{1}{\gamma_2-2}-\frac{1}{\gamma_1-2}\big ]}$ and $\xi_n=0.8$. 
The neutron spectrum can be expressed in terms of $E_\gamma^{iso}$ and the equipartition parameters $\epsilon_e$, $\epsilon_B$ as follows,
\begin{equation}
\epsilon_n^2\frac{dN_n}{d\epsilon_n}=\frac{f_n}{2\kappa}\frac{1-\epsilon_e-\epsilon_B}{\epsilon_e}
{E_{\gamma}^{iso}}.
\label{totnunu}
\end{equation}
We have assumed the probability of production of neutrons in resonant $p\gamma$  interactions to be half. The energy flux of antineutrinos of energy $\epsilon_{\bar{\nu},n}$  can be estimated with the neutron flux $({dN_n}/{d\epsilon_n})$ 
at the source rest frame \cite{anc2} as,
\begin{equation}
\epsilon_{\bar{\nu},n}^2\frac{dN_{\bar{\nu}}}{d\epsilon_{\bar{\nu},n}}(\epsilon_{\bar{\nu},n})
=\left[\int\limits_{\frac{m_n\,\epsilon_{\bar{\nu},n}}{2\,\epsilon_0}}^{\epsilon_{n,max}}
\frac{d\epsilon_n}{\epsilon_n}\,\frac{dN_n}{d\epsilon_n}\left(1-e^{-\frac{D_sm_n}{\epsilon_n\,\overline{\tau}_n}}\right)\frac{m_n}{2\,\epsilon_0} \,\right] \times \epsilon_{\bar{\nu},n}^2.
\label{nu}
\end{equation}
$\epsilon_0$ is the mean energy of an antineutrino in the neutron rest frame.
 The bracketed term is the decay probability for a neutron with energy $\epsilon_n$ travelling a distance $D_s$. However if the source is at a large distance then all the neutrons are expected to decay before reaching earth.
 The maximum energy of the secondary neutrons is $\epsilon_{n,max}=\xi_n\epsilon_{p,max}$. The maximum energy of the antineutrinos from neutron decay is $Q/m_n$ times the maximum energy of the decaying neutrons, where $Q=m_n-m_p-m_e$ and $m_n$, $m_p$, $m_e$ are the masses of neutron, proton and electron respectively. 
 
\subsection{Neutrinos from kaon decay}
In $p\gamma$ interactions charged kaons ($K^+$) are produced through the following interactions $p\gamma \rightarrow K^+ \Lambda^0$ and $p\gamma \rightarrow K^+ \Sigma^0$. 
Although the cross-sections of these interactions are lower compared to photo-pion production \cite{gla}, at very high energy the total neutrino flux produced through 
kaon decay becomes higher than that from photo-pion decay \cite{asan,wal1}. The cross-section for $K^+$ and $\Lambda^0$ production peaks at $\sigma_{\Lambda^0,K^+} = 2 \times 10^{-30}$ $cm^2$ 
for photon energy $\epsilon_0^{K^+} = 1.3 $ GeV in the proton rest frame and it has a full width at half maximum of $\delta\epsilon_{K^+}$= 0.9 GeV. $K^+$ and $\Sigma^0$ are
produced with the highest cross-section of $\sigma_{\Sigma^0,K^+} = 2.4 \times 10^{-30}$ $cm^2$ at $\epsilon_0^{K^+} = 1.45 $ GeV, and $\delta\epsilon_{K^+}$= 0.8 GeV in this case.  
The average fraction of energy lost by a proton in producing a $K^+$, 
is  $\xi_{K^+}= \frac{1}{2}\big[1-\frac{m^2_B(\Lambda^0 \, or \, \Sigma^0)-m_{K^+}^2}{S}\big]$. S is the invariance of the square of the total four-momentum
of the p$\gamma$ system and $m_\Lambda^0$, $m_\Sigma^0$,
$m_{K^+}$ are the masses of $\Lambda^0$, $\Sigma^0$ and $K^+$ respectively.
The threshold energy of photon in 
proton rest frame for kaon and $\Lambda^0$ production is $\epsilon_{th,{K^+}}$  $\sim$ 455 MeV. Fig.1 shows the inelasticity for photo-pion and $K^+$ with $\Lambda^0$ production in $p\gamma$ interactions plotted against the photon energy in the proton rest frame. 
 \begin{figure}
  \vspace{5mm}
  \centering
  \includegraphics[width=3.in]{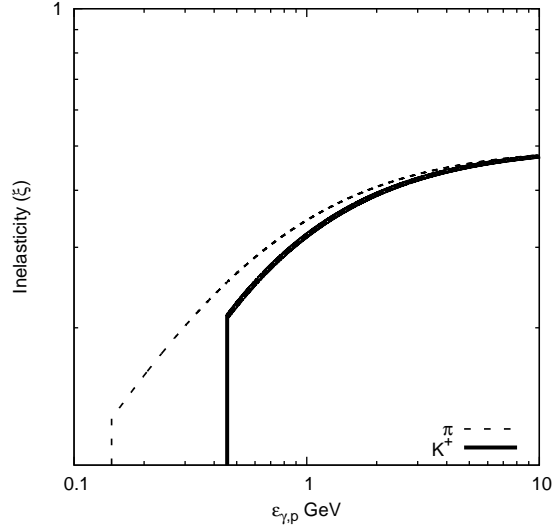}
  \caption{Inelasticity of meson production from proton photon interaction as a function of photon energy in proton rest frame.}
  \label{inel_fig}
 \end{figure}
The cross-sections are measured upto $\epsilon_{\gamma,p}=2.6$GeV in the paper by \cite{gla}. The fractional energy transferred from shock accelerated protons to kaons has been calculated in the same way as for photo-pions.
\\
In the case of resonant production of $K^+$ with $\Lambda^0$, the fractional energy loss of a proton of energy $\epsilon_p$ is the ratio of kaon production and dynamical time scale of the GRB,
\begin{equation}
f_{K^+} (\epsilon_p)= f_0^{K^+} \left\{\begin{array}{l@{\quad\quad}l}
\frac{0.7^{\gamma_1-1}}{\gamma_1+1}\left(\frac{\epsilon_p}{\epsilon_{p,\Lambda^0}^b}\right)^{\gamma_1-1}-2.03\left( \frac{1}{\gamma_1+1}- \frac{1}{\gamma_2+1}\right)\left(\frac{\epsilon_p}{\epsilon_{p,\Lambda^0}^b}\right)^{-2}& {\epsilon_p>\epsilon_{p,\Lambda^0}^b}\\
\frac{0.7^{\gamma_2-1}}{\gamma_2+1}\left(\frac{\epsilon_p}{\epsilon_{p,\Lambda^0}^b}\right)^{\gamma_2-1}&
{\epsilon_p < \epsilon_{p,\Lambda^0}^b}
\end{array}
\right.
\label{fK+}
\end{equation}
In the above equation $f_0^{K^+}=\xi_{K^+}\frac{0.019L_{\gamma,51}}{\Gamma_{300}^4 \,
t_{v,-3}{\epsilon_{\gamma,MeV}^b}}\frac{1}{\big
[\frac{1}{\gamma_2-2}-\frac{1}{\gamma_1-2}\big ]}$ with the mean value of 
$\xi_{K^+} \sim 0.2$ \cite{asan}.
The protons of energy $\epsilon_{p,\Lambda^0}^b$ satisfy the threshold energy condition for production of $K^+$ and $\Lambda^0$ in interactions with low energy photons of energy $\epsilon_{\gamma,MeV}^b$.
The expression for $\epsilon_{p,\Lambda^0}^b$ is as given below
\begin{equation}
\epsilon_{p,\Lambda^0}^b=3.84\times10^{7}\Gamma_{300}^2(\epsilon_{\gamma,MeV}^b)^{-1}\, {\rm GeV}.
\end{equation}

If we consider $p\gamma$ interactions with multiparticles at the final states,
 $p\gamma \rightarrow K^+ \Lambda^0 \pi^0$, then the cross-section for $K^+$ production is $\sigma_{K^+,mult} \approx 0.5\times 10^{-30}$ $cm^2$ as measured by \cite{law} for photon energy upto 2.6GeV in proton rest frame.
In this case we have assumed $20\%$ of a proton's energy goes to $K^+$.

\begin{equation}
f_{K^+}^{mult} (\epsilon_p)= f_{0,mult}^{K^+} \left\{\begin{array}{l@{\quad\quad}l}
\frac{1}{(\gamma_1+1)(\gamma_1-1)}\left(\frac{\epsilon_p}{\epsilon_{p,mult}^b}\right)^{\gamma_1-1}
+0.5(\frac{1}{\gamma_1+1}-\frac{1}{\gamma_2+1})\left(\frac{\epsilon_p}{\epsilon_{p,mult}^b}\right)^{-2}+{\cal O}(\gamma_1,\gamma_2) 
\hspace{0.1cm}{\epsilon_{p}> \epsilon_{p,mult}^b}\\
\frac{1}{(\gamma_2+1)(\gamma_2-1)}\left(\frac{\epsilon_p}{\epsilon_{p,mult}^b}\right)^{\gamma_2-1} \hspace{2.cm} {\epsilon_p<\epsilon_{p,mult}^b}. 
\end{array}
\right.
\label{fK}
\end{equation}
$f_{0,mult}^{K^+}=\xi_{K^+,mult}\frac{0.0013L_{\gamma,51}}{\Gamma_{300}^4 \,
t_{v,-3}{\epsilon_{\gamma,MeV}^b}}\frac{1}{\big
[\frac{1}{\gamma_2-2}-\frac{1}{\gamma_1-2}\big ]}$ 
, $O(\gamma_1,\gamma_2)=\frac{1}{(\gamma_2+1)(\gamma_2-1)}-\frac{1}{(\gamma_1+1)(\gamma_1-1)}-\frac{0.5(\gamma_2-\gamma_1)}{(\gamma_1+1)(\gamma_2+1)}$, and,

\begin{equation}
\epsilon_{p,mult}^b=4.9\times10^{7}\Gamma_{300}^2(\epsilon_{\gamma,MeV}^b)^{-1}\, {\rm GeV}.
\end{equation}
$K^+$ decay to secondary neutrinos by the following channels, 
$K^+ \rightarrow \mu^+\nu_{\mu}(63\%)$, $\pi^+\pi^0(21\%)$, 
$\pi^+\pi^+\pi^-(6\%)$, $\pi^0 e^+\nu_e(5\%)$, $\pi^0 \mu^+\nu_\mu(3\%)$ and 
$\pi^+\pi^0\pi^0(2\%)$.   
Due to their heavier mass $K^+$ cools at higher energy compared to muon and pion. The synchrotron cooling break energy in the kaon spectrum is at
\begin{equation}
 \epsilon_K^s= 2.2\times10^{9}\epsilon_e^{1/2}\epsilon_B^{-1/2}
L_{\gamma,51}^{-1/2}\Gamma_{300}^4t_{v,-3} \, {\rm GeV}.
\end{equation}
It is derived by comparing the decay and cooling time scales of kaons.
 Although the cross-section of kaon production is much less than that of 
pion production, the neutrino flux from kaon channel exceeds the flux from pion channel at very high energy due to the slower rate of cooling of kaons. Similarly one can do the calculation for $K^+$ and $\Sigma^0$ production in $p\gamma$ interactions. The neutrino flux from this channel will be added to the previous channel (eqn. \ref{totnuKa}). \\
 Neutral kaons are produced in $p\gamma$ interactions with a cross-section $\sigma_{K^0,\Sigma} \approx 0.6 \times 10^{-30}$ $cm^2$ \cite{law} 
 at the peak energy ${\epsilon_0}^{K^0}=1.45$ GeV and width $\delta\epsilon_{K^0}$ = 0.7 GeV. Half of the neutral kaons are assumed to be long lived kaons ($K_L^0$). $K^0$ can be produced in $p\gamma$ interactions with multiparticle final states ($p\gamma \rightarrow K^0_S \Lambda^0 \pi^+$, $ K^0_L \Lambda^0 \pi^+$ and $K^0_S \Sigma^+ \pi^0$). The cross-sections of these interactions are measured as $0.5 \times 10^{-30}cm^2$, $0.5 \times 10^{-30}cm^2$ and $0.2 \times 10^{-30}cm^2$ respectively \cite{law}.
 $K_L^0$ decays through the following channels $\pi^+e^-\bar{\nu_e}(39\%)$, $\pi^+\mu^-\bar{\nu_\mu}(27\%)$, $\pi^0 \pi^0 \pi^0(21\%)$, and $\pi^+ \pi^- \pi^0(13\%)$. $K_S^0$ decays to two charged pions through this channel $\pi^+\pi^-(69\%)$. The pions finally decay to neutrinos and antineutrinos. The total neutrino flux from $K_L^0$ and $K_S^0$ can be calculated in the same way as $K^+$.\\
 The energy flux of neutrinos with energy $\epsilon_{\nu,K}$ from kaon decay in GRB source rest frame is,
\begin{eqnarray}
 \epsilon_{\nu,K}^2\frac{dN_{\nu}(\epsilon_{\nu,K})}{d\epsilon_{\nu,K}} &\approx& \frac{\, f_{\nu,K} \,f_{K}}
{\kappa}\frac{(1-\epsilon_e-\epsilon_B)}{\epsilon_e}E_{\gamma}^{iso}\left\{\begin{array}{l} 1 \hspace{1.cm} \epsilon_{\nu,K}<\epsilon_{\nu,K}^{s}\\ 
 \left(\frac{\epsilon_{\nu,K}}{\epsilon_{\nu,K}^{s}}\right)^{-2} \hspace{0.2cm} \epsilon_{\nu,K}>\epsilon_{\nu,K}^{s}.
\end{array}
 \right.
\label{totnuKa}
\end{eqnarray}
 $f_{\nu,K}$ is the fractional energy of a kaon transferred to each neutrino/antineutrino produced in kaon decay.
The observed total neutrino flux with energy $\epsilon_{\nu}^{ob}$ on earth is,
\begin{equation}
\frac{dN_{\nu}^{ob}(\epsilon_{\nu}^{ob})}{d\epsilon_{\nu}^{ob}}=\frac{dN_{\nu}(\epsilon_{\nu})}{d\epsilon_{\nu}}\frac{1+z}{4\pi
D_s^2} 
\end{equation}
where $z$ is the redshift of the GRB.
\subsection{Results and Discussions}
 Our calculations are based on the standard internal shock model of GRBs. In internal shocks the shock radius $r_d$ is related to the bulk Lorentz factor $\Gamma$ and variability time $t_v$, $r_d=\Gamma^2 c t_v$, where $c$ is the speed of light. We have not assumed any relation among the GRB parameters $\Gamma$ and isotropic energy \cite{Liang,Lv,ghr1} or peak luminosity and observed break energy in the low energy photon spectrum \cite{ghr1}.
The values of the parameters of GRBs have been varied to show that the importance of the different channels may vary depending on the luminosities, Lorentz factors, variability times and other parameters of GRBs. The redshifts of the GRBs have been denoted by $z$.  
We have assumed the GRBs have durations of prompt emission 5 sec. Our results for various set of values of the
 parameters of the GRBs are shown in Figure 2 to Figure 6. The following values 
of the parameters have been used,

 Figure 2. $\gamma_1=1$, $\gamma_2=2.2$, $z=1.8$, $L_\gamma=10^{53}$ erg/sec, $\Gamma=600$, $t_v=20$ msec, $\epsilon_\gamma^b=0.5 MeV$, $\epsilon_B$=$\epsilon_e=0.3$, $f_{\pi}^0=0.09$ and $r_d=2.16 \times 10^{14}$ cm, $B_c=1.54 \times 10^{5}$ G.\\

 Figure 3. $\gamma_1=1.2$, $\gamma_2=2.5$, $z=1.2$, $L_\gamma=10^{52}$ erg/sec, $\Gamma=600$, $t_v=20$ msec, $\epsilon_\gamma^b=0.5 MeV$, $\epsilon_B=0.6$, $\epsilon_e=0.06$,  $f_{\pi}^0=0.14$ and $r_d=2.16 \times 10^{14}$ cm, $B_c=1.219 \times 10^{3}$ G.\\ 

Figure 4. $\gamma_1=1.8$, $\gamma_2=2.01$, $z=1$, $L_\gamma=5\times10^{51}$ erg/sec, $\Gamma=130$, $t_v=25$ msec, $\epsilon_\gamma^b=0.5 MeV$, $\epsilon_B$=$\epsilon_e=0.3$, $f_{\pi}^0=0.16$ and $r_d=1.26 \times 10^{13}$ cm, $B_c=0.222 \times 10^{5}$ G.\\

 Figure 5. $\gamma_1=1.2$, $\gamma_2=2.2$, $z=0.8$, $L_\gamma=10^{52}$ erg/sec, $\Gamma=1000$, $t_v=10$ msec, $\epsilon_\gamma^b=0.5 MeV$, $\epsilon_B$=$\epsilon_e=0.1$, $f_{\pi}^0=0.054$ and $r_d=3 \times 10^{14}$ cm, $B_c=4.303 \times 10^{3}$ G.\\

Figure 6. $\gamma_1=1$, $\gamma_2=2.2$, $z=1$, $L_\gamma=10^{51}$ erg/sec, $\Gamma=600$, $t_v=2$ msec, $\epsilon_\gamma^b=0.5 MeV$, $\epsilon_B$=$\epsilon_e=0.3$, $f_{\pi}^0=0.009$ and $r_d=2.16 \times 10^{14}$ cm, $B_c=1.543 \times 10^{5}$ G.\\

The total neutrino fluxes from the decay of muons, pions and neutrons are shown in Figure ``a''. The total neutrino flux from the decay of muons and pions, where the pions were produced in resonant interactions of shock accelerated protons with low energy photons, is shown by solid blue line. Red solid line is for the total neutrino flux from the decay of neutrons, where the neutrons were produced in resonant $p\gamma$ interactions. 
At low energy the total neutrino flux from neutron decay can exceed the sum of the total neutrino fluxes from muon and pion decay as shown in 
Figure 3[a] and 3[c]. At very high energy the neutron decay channel may also become important as shown in Figure 4[a] and 4[c].

 The total neutrino fluxes from the decay of charged and neutral kaons are shown in Figure ``b''. Charged kaons ($K^+$) are produced in resonant interactions of shock accelerated protons with the low energy photons. The total neutrino fluxes produced in the decay of charged kaons are shown by solid orange line for the resonant production of $K^+$. The dashed orange line is for the total neutrino flux from the decay of $K^+$ where the charged kaons were produced in $p\gamma$ interactions with multiparticle final states. Similarly solid, dashed cyan and magenta lines represent the total neutrino fluxes from decaying long lived ($K^0_L$) and short lived ($K^0_S$) neutral kaons for the above mentioned cases. Green solid line shows the total neutrino flux from the charged and neutral kaons after summing up all the different components mentioned above. The total neutrino flux produced from kaon decay shows distinct peak at the highest energy.\\
Figure ``c'' represents the total fluxes of neutrinos from the decay of different secondary particles ($\mu^+$, $\pi^+$ (blue), neutron (red) and kaon (green)). After adding up the contributions from all the decaying secondaries the total neutrino flux has been shown by black solid line.\\
 The peak in the neutrino spectrum near PeV energy is due to protons of energy 200 PeV.

The flavour ratio of neutrinos can be an useful observable to reveal the underlying physical phenomena in many processes \cite{wax3,pak,pun}. 
Kashti $\&$ Waxman \cite{wax3} explored the energy dependence of the neutrino flavour ratios produced in $pp$ and $p\gamma$ interactions. Energetic pions and muons cool by emitting radiations in the internal magnetic field of the sources. As muons cool faster compared to pions muon decay neutrino flux is suppressed at high energy. This leads to a flux ratio of neutrinos and antineutrinos $\phi_{\nu_{e}}:\phi_{\nu_{\mu}}:\phi_{\nu_{\tau}}=1:1.8:1.8$ expected on earth instead of $1:1:1$. This flux ratio corresponds to $R=\phi_{\mu}/(\phi_{e}+\phi_{\tau})=
0.64$ on earth.

In the context of high energy neutrinos from GRBs the flavour ratio $R=\phi_{\mu}/(\phi_{e}+\phi_{\tau})$ expected on earth has been calculated in \cite{wal2}.
In our Figure ``d'' we have plotted the flavour ratio R expected on earth 
from our calculations. Our plots can be compared with those obtained by \cite{wal2}. Due to the contributions from neutron and kaon decay neutrinos and antineutrinos the plots of flavour ratio $R$ in our work show different features.  In Figure 4[d] the sudden dips are due to the importance of the neutron decay channel (near $10^7$ GeV) and kaon decay channels (near $10^9$ GeV) compared to the pion decay channel. In rest of the plots (Figure 2[d], 3[d], 5[d], 6[d]) showing the energy dependence of R, the first bump is due to the decrease in neutron decay antineutrinos above $10^5 GeV$ and the second bump is due to muon cooling.
The step at higher energy after the second bump is due to the importance of the kaon decay channel compared to the pion decay channel. 
 In the paper by Baerwald et al. \cite{wal2} below 100 TeV
 the flavour ratio $R$ is between 0.36 and 0.5 after including variations in values of all the GRB parameters. 
Above 10 PeV they have $R=0.64$ due to muon damping assuming at the source the
 flux ratios are $\phi_{\nu_e}:\phi_{{\nu}_{\mu}}:\phi_{\nu_{\tau}}=0:1:0$. 
We have not made such assumptions in our calculations. We have included muon cooling at high energy with the neutrino flux given in eqn.(\ref{totnupi}). We 
have also included the kaon decay neutrinos in our calculation of $R$, as a result our calculated R does not remain constant at high energy. We have used 
$sin^2{2\theta_{13}}=0.1$, where $\theta_{13}$ is the neutrino mixing angle 
measured by DOUBLE CHOOZ \cite{chooz}, DAYA-BAY \cite{daya} and RENO collaborations \cite{reno}.
In our Figure 2[d] at 5 TeV the flavour ratio is $R=0.41$, however in Figure 3[d] neutron decay channel is more important at low energy and $R=0.36$. These results can be compared with a pure neutron beam source where $R=0.28$ and for a pure pion beam source this ratio is 0.5.
In Figure 4[d] upto 1 PeV the flavour ratio $R$ is $0.47$ then it goes down to $0.37$ at 7.5 PeV as the neutron decay channel dominates at this energy.
In Figure 5[d] and 6[d] at 1 TeV the flavour ratio is $R=0.41$.  
In Figure 2[d], 3[d], 5[d] and 6[d] near 1 PeV muon and pion decay neutrino flux becomes higher than the neutron decay antineutrino flux and $R$ increases to $0.48$. In Figure 2[d] due to muon cooling R increases to 0.54 near 2 EeV and then it attains the highest value of 0.57 at 10 EeV due to the kaon decay channels. In Figure 3[d] near 1 EeV muon cooling starts, at the same time kaon channel becomes important. $R$ is 0.51 at 1 EeV and near 30 EeV $R$ is 0.57. In Figure 4[d], 5[d] and 6[d] this highest value of R is attained above 1 EeV, 80 EeV and 10 EeV respectively. 
At 1 EeV energy our $R$ values are almost $20\%$ to $30\%$ lower than the value of 0.64 obtained by Baerwald et al. \cite{wal2}. At the highest energy our $R$ values of 0.57 are $12\%$ lower than the constant value of 0.64 in Baerwald et al. \cite{wal2}. One may notice from our figures that the muon cooling and the importance of the kaon decay channels together determine the ratio $R$ at high energy.
 
It is interesting to note that the flavour ratio $R$ carries the information of the origin of the neutrinos and the antineutrinos.

We have shown that the different decay processes may contribute differently depending on the values of the parameters of the GRBs.
The energy of the neutrinos above which the neutron decay channel will dominate over the pion decay channel can be derived by equating the 
fluxes from the two channels with $\epsilon_{\bar{\nu},n}$=$\epsilon_{{\nu},\pi}$. We compare the total neutrino spectrum from pion decay above the synchrotron cooling break energy (eqn. \ref{totnupi}) with the same from 
neutron decay (eqn. \ref{nu}) assuming all the neutrons have decayed during propagation.

\begin{equation}
{\epsilon_{\nu}}^2 \left[1-\left(\frac{10^{10.33} \, GeV \epsilon_{n,10.33}^{max}}{{m_n\epsilon_{\bar{\nu},n}}/{2\epsilon_0}}\right)^{\gamma_1-3}\right] \geq  
1.96 \times 10^{13} (3-\gamma_1) \left(\frac{f_{0,0.2}^\pi}{f_{0,0.8}^n}\right)  
\left(\frac{\epsilon_{n,7.01}^b}{\epsilon_{\nu,\pi,5.81}^b}\right)^{\gamma_1-1} \left(\frac{m_n}{2\epsilon_0}\right)^{2-\gamma_1}
{{\epsilon_{\nu,\mu,6.4}^s}^2}  \, GeV^2
\label{compinu}
\end{equation}
The energies $\epsilon_{n,10.33}^{max}=\epsilon_{n}^{max} /(10^{10.33} GeV) $, 
$\epsilon_{\nu,\mu,6.4}^s=\epsilon_{\nu,\mu}^s/(10^{6.4} \, GeV) $, $\epsilon_{n,7.01}^b=\epsilon_{n}^b /(10^{7.01} \,GeV)$ and $\epsilon_{\nu,\pi,5.81}^b= \epsilon_{\nu,\pi}^b/(10^{5.81} \, GeV) $ are calculated for the following set of GRB parameters $\gamma_1=1.2$, $\gamma_2=2.3$, $L_\gamma=10^{51}$ erg/sec, $\Gamma=300$, $t_v=1.$ msec, $\epsilon_\gamma^b=1 MeV$.
Solving this equation numerically one can find the energy at which neutrinos from pion decay channels will be suppressed below the neutron decay channel. Similar to eqn. (\ref{compinu}) one can get the inequality relation for the energy above which the neutrino flux from the kaon decay channel (eqn. \ref{totnuKa}) will dominate over the pion decay channel (eqn.\ref{totnupi})
\begin{equation}
{\epsilon_{\nu}}^2 \geq 4.09 \times 10^{14} \, (1.54)^{\gamma_1-1}f_{\nu,K} \left(\frac{f_{0,0.2}^\pi}{f_{0,0.001}^{k^+}}\right)  
\left(\frac{\epsilon_{p,\Lambda^0,7.59}^b}{\epsilon_{p,\Delta,7.11}^b}\right)^{\gamma_1-1} {{\epsilon_{\nu,\mu,6.4}^s}^2} \, GeV^2 
\label{com}
\end{equation}
where $f_{\nu,k}$ is the fractional energy of a kaon going to each neutrino/antineutrino, and $\epsilon_{p,\Lambda^0,7.59}^b= \epsilon_{p,\Lambda^0}^b/(10^{7.59} \, GeV)$, $\epsilon_{p,\Delta,7.11}^b= \epsilon_{p,\Delta}^b/(10^{7.11} \,GeV)$, same set of values of the GRB parameters has been used here.  
These inequalities depend on the GRB parameters, $L_\gamma$, $\Gamma$, $t_v$ and $\gamma_1$, $\gamma_2$, break energy in the photon spectrum. A small change in  values of any of the parameters will lower one channel with respect to another and change the shape of the total neutrino spectrum from a GRB.
The effect of stochastic reacceleration on photon and neutrino spectra from GRBs has been studied for photospheric radii of GRB fireballs and high Lorentz boost factors in \cite{murase4}. Inclusion of stochastic reacceleration may shift the peak in the total neutrino fluences to a higher energy as given in their Fig.7. In that case the contribution of the neutron decay channel to the total neutrino spectrum may still remain important at low energy.
We have discussed about the limits obtained by IceCube experiment in the Introduction of this paper. We compare our results with the recent IceCube limit 
\cite{abbasi4} for GRB neutrinos with 40 and 59 strings. The details of their calculations and the values of the parameters they have used are given in \cite{abbasi0,abbasi1,abbasi4}. 
   IceCube collaboration observed 117 GRBs with 40-string, two additional GRBs included from test runs before the official start of 59-string and 181 GRBs with 59-string. The sum of the neutrino energy fluences from all the 353 GRBs observed is shown in their Figure.1. of \cite{abbasi4} with the lower most black solid line. We have shown in our Figure [c] the same with solid brown line \cite{abbasi4}. Our calculated GRB neutrino fluences from 353 GRBs for each set of parameter values are shown with brown dashed lines in Figure 2[c]-6[c]. Our choice of values of the GRB parameters give neutrino fluences below the current limit set by the IceCube expriment.    
\section{Conclusion}
Gamma ray bursts have been speculated to produce high energy particles like cosmic rays, gamma rays and neutrinos. Ultrahigh energy muons, pions, neutrons and kaons could be produced in GRBs due to p$\gamma$ interactions.
These particles subsequently decay to produce very high energy neutrinos and antineutrinos. Charged particles lose their energy due to synchrotron cooling in the magnetic field before decaying to lighter particles. Kaons suffer synchrotron loss at higher energy compared to muons and pions. As a result at higher energy the kaon decay channel of neutrino production dominates over the muon and pion decay channels.We have included all the decay channels of charged ($K^+$) and neutral kaons ($K^0_L$ and $K^0_S$) to study this effect.
The total neutrino flux from neutron decay channel may be more important than that from muon and pion decay channels depending on the values of the parameters of GRBs. The flavour ratios of neutrinos show distinct features depending on the values of the GRB parameters and the dominance of different decay processes at different energies. IceCube has set limit on the GRB neutrino flux. Future detectors with higher sensitivities will be able to reveal the role of GRBs as a UHECR accelerator.  

\section{Acknowledgement}
RM wishes to thank RRI, Bangalore for hospitality where a major part of this work was done. We thank the referee for useful comments and clarifications.

\clearpage

\begin{figure}
  \begin{center}
    \begin{tabular}{cc}
      \resizebox{80mm}{!}{\includegraphics{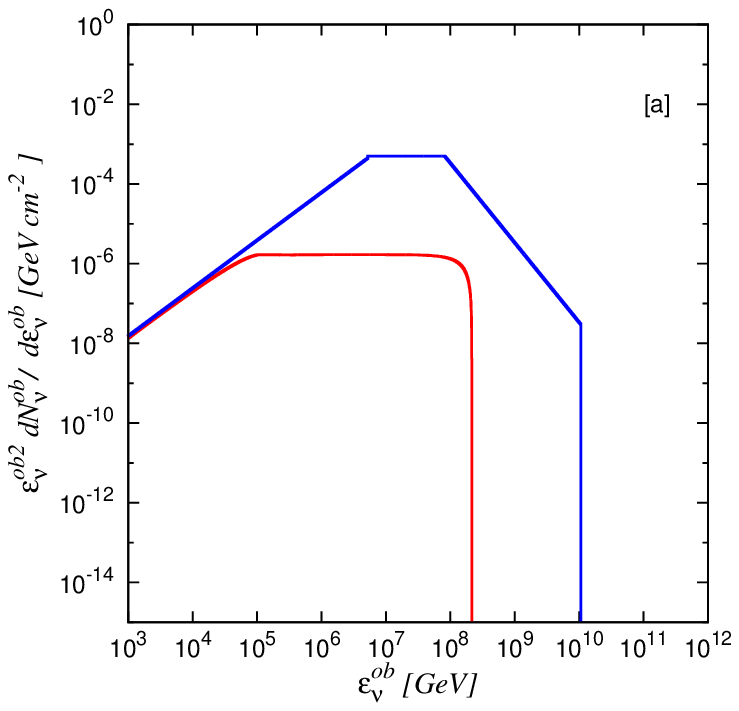}} &
      \resizebox{80mm}{!}{\includegraphics{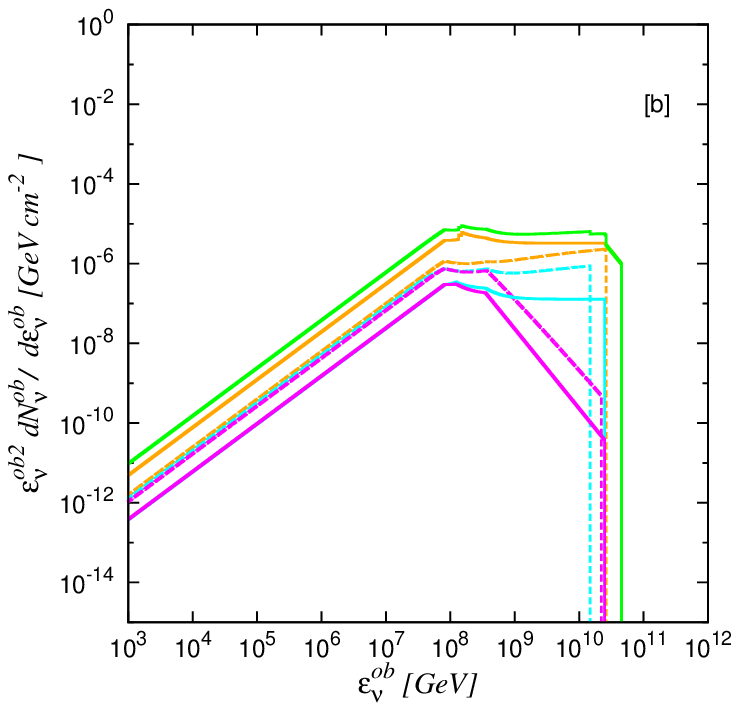}} \\
      \resizebox{80mm}{!}{\includegraphics{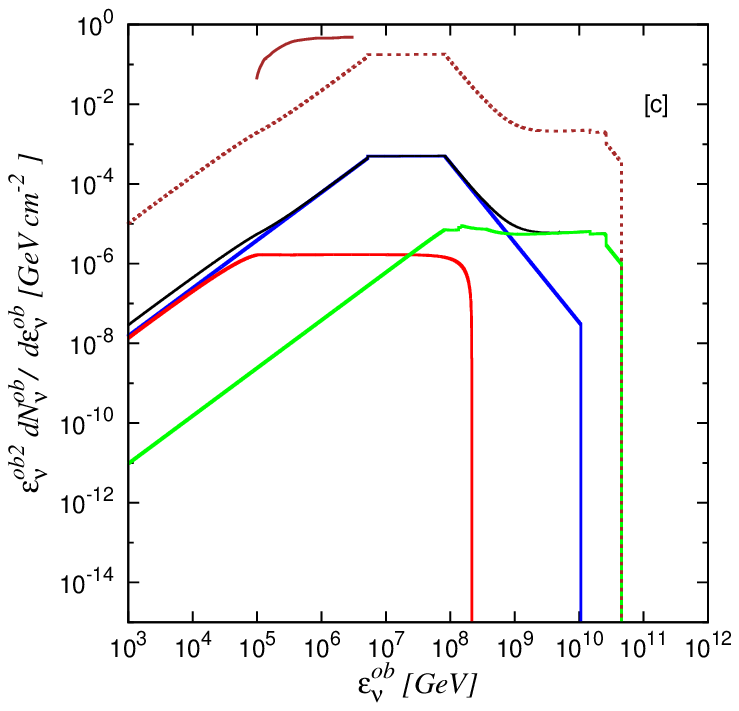}} &
      \resizebox{80mm}{!}{\includegraphics{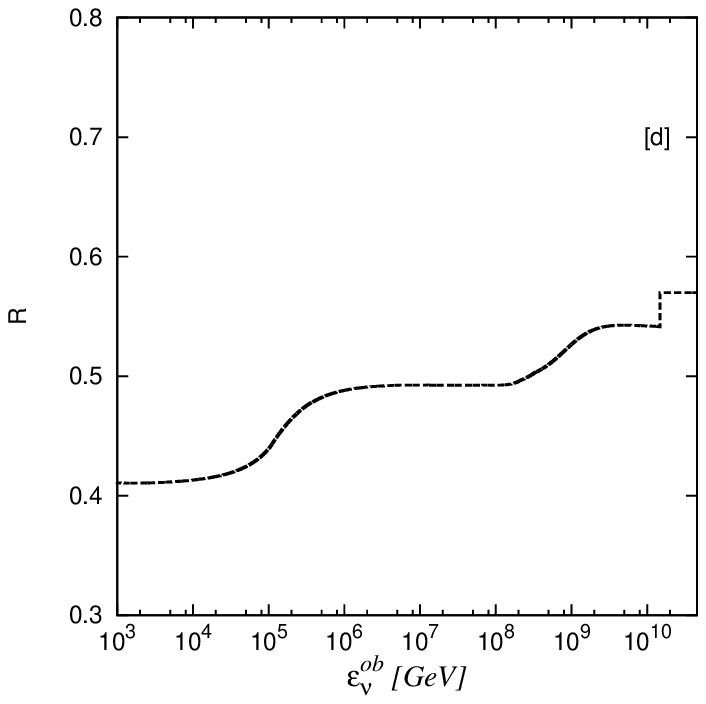}} \\
    \end{tabular}
    \caption{ $\gamma_1=1$, $\gamma_2=2.2$, $z=1.8$, $L_{\gamma}=10^{53}$erg/sec, $\Gamma=600$, $t_v=20$msec, $\epsilon^b_{\gamma}=0.5$MeV, $\epsilon_B=\epsilon_e=0.3$.
[a] Blue solid: resonant pion, muon, red solid: resonant neutron decay neutrinos; [b] orange solid, dashed: resonant, multiparticle final state $K^{+}$ decay neutrinos, cyan solid, dashed for $K^0_L$, magenta solid, dashed for $K^0_S$, green solid: total neutrinos from kaon decay; [c] blue: total neutrinos from muon and pion decay, red: neutron decay, green: kaon decay; black: total neutrinos from decay of all particles; brown solid: IceCube-40+59 string limit, brown dashed: our aggregate of 353 GRBs [d] flavour ratio $R$ as discussed in the text}
    \label{test4}
  \end{center}
\end{figure}

\clearpage

\begin{figure}
  \begin{center}
    \begin{tabular}{cc}
      \resizebox{80mm}{!}{\includegraphics{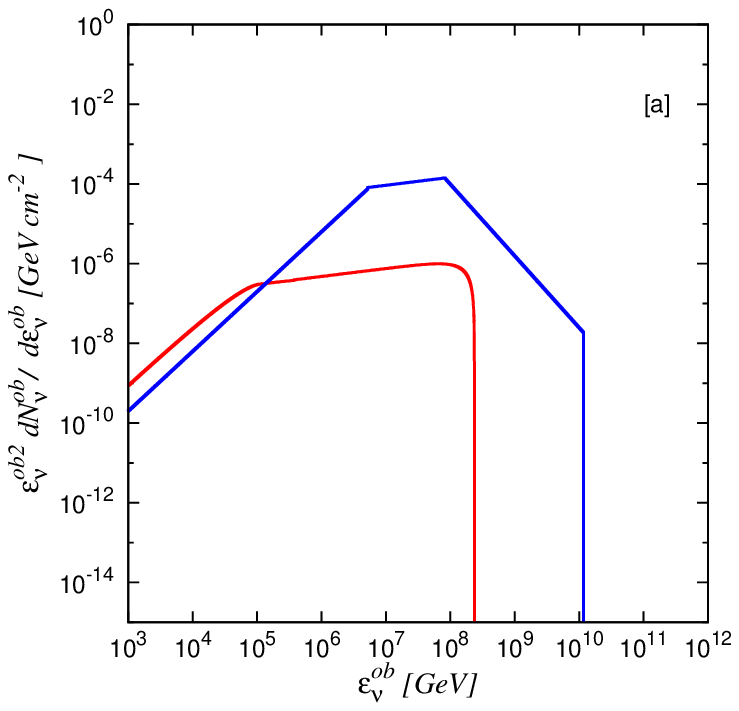}} &
      \resizebox{80mm}{!}{\includegraphics{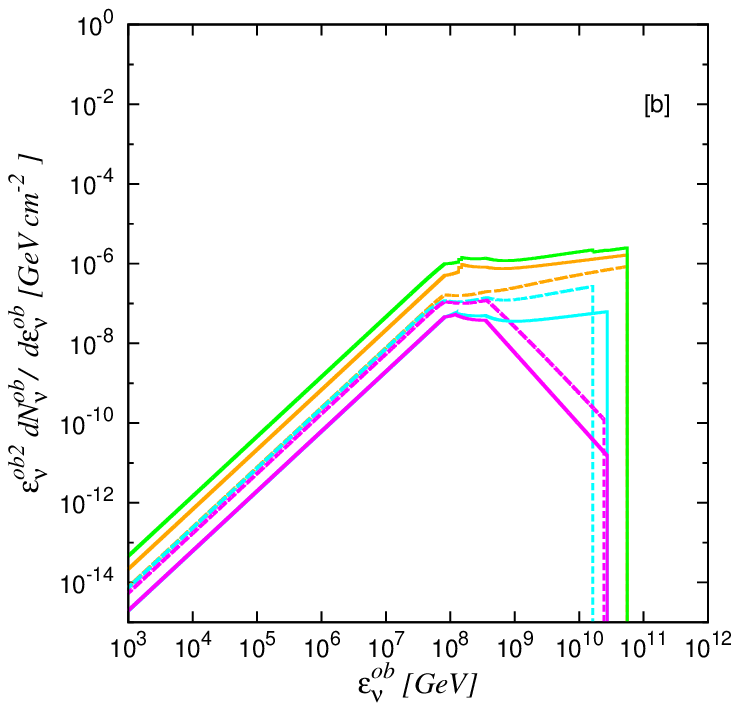}} \\
      \resizebox{80mm}{!}{\includegraphics{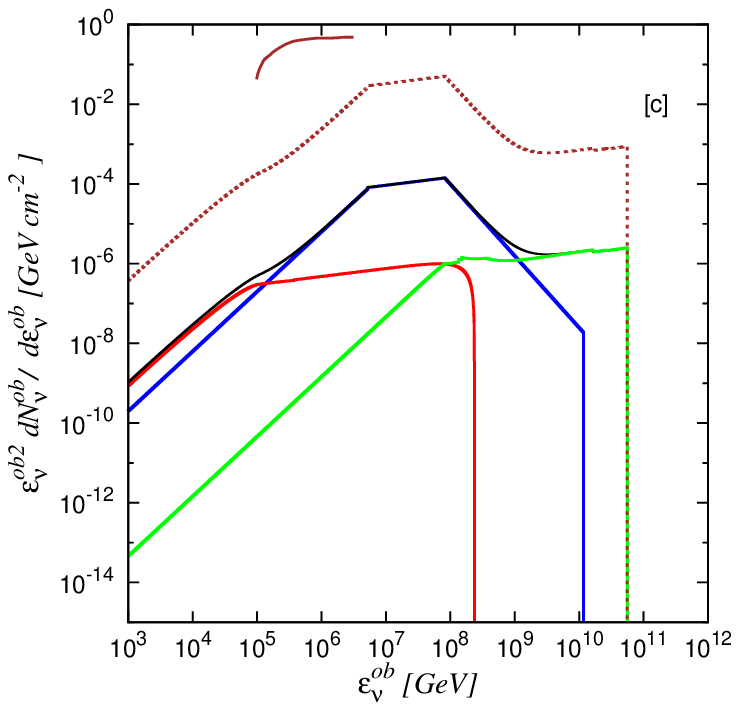}} &
      \resizebox{80mm}{!}{\includegraphics{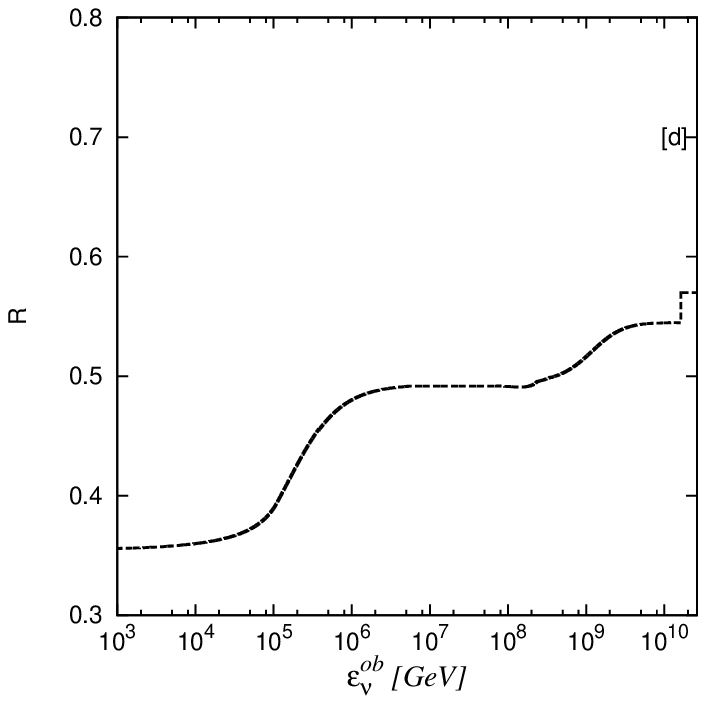}} \\
    \end{tabular}
    \caption{$\gamma_1=1.2$, $\gamma_2=2.5$, $z=1.2$, $L_{\gamma}=10^{52}$erg/sec, $\Gamma=600$, $t_v=20$msec, $\epsilon^b_{\gamma}=0.5$MeV, $\epsilon_B=0.6$, $\epsilon_e=0.06$. Line styles same as in Figure 2.}
    \label{test4}
  \end{center}
\end{figure}

\clearpage

\begin{figure}
  \begin{center}
    \begin{tabular}{cc}
      \resizebox{80mm}{!}{\includegraphics{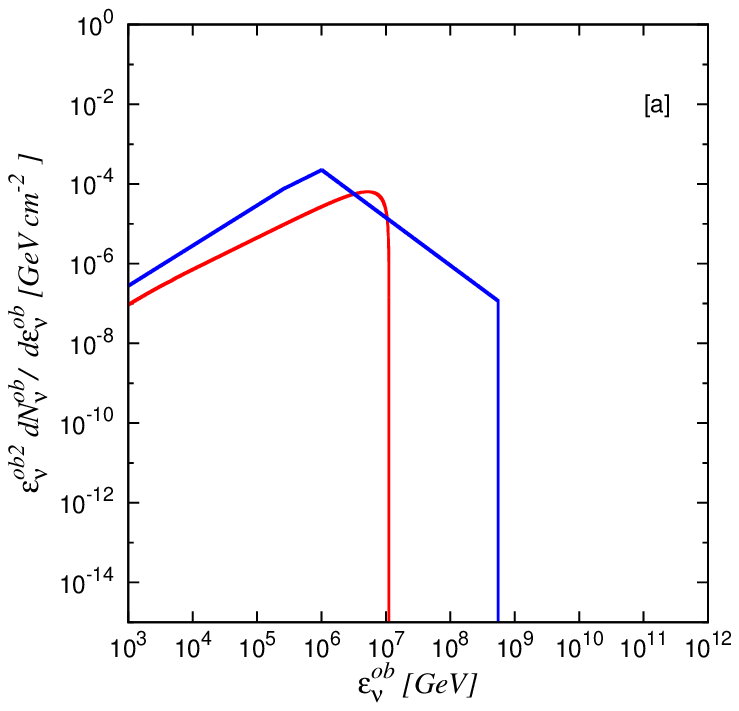}} &
      \resizebox{80mm}{!}{\includegraphics{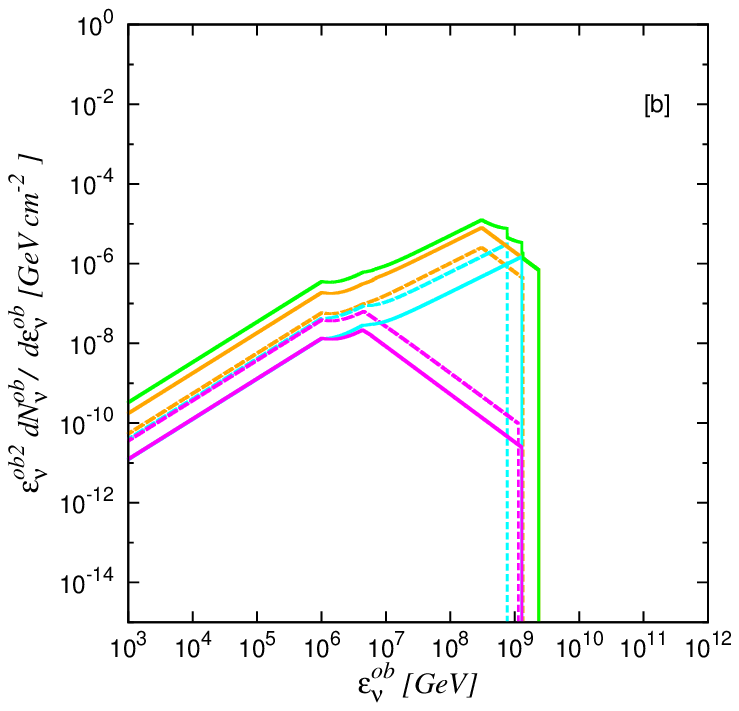}} \\
      \resizebox{80mm}{!}{\includegraphics{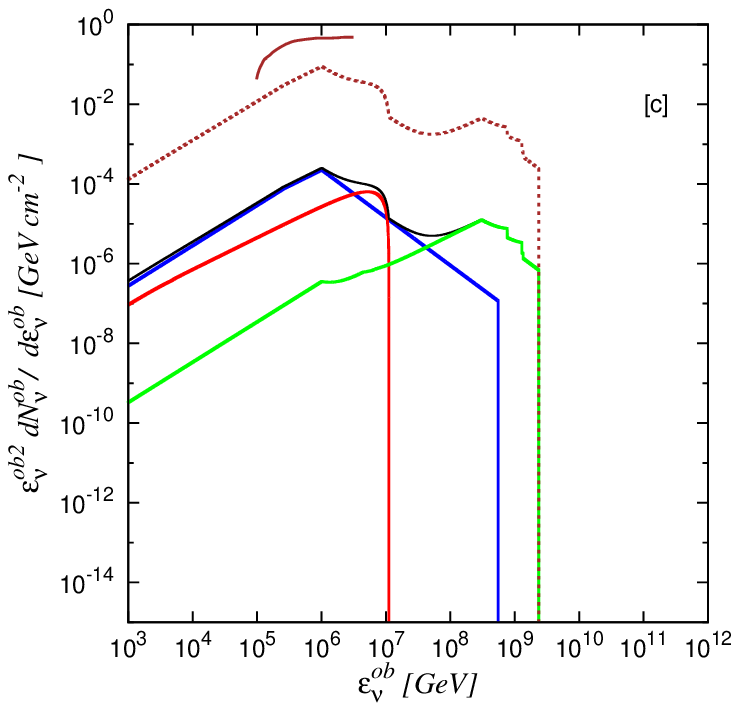}} &
      \resizebox{80mm}{!}{\includegraphics{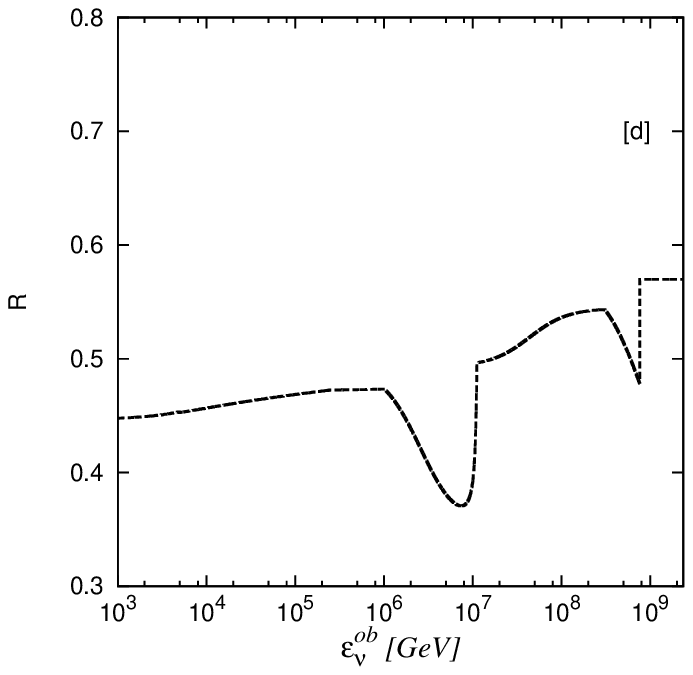}} \\
    \end{tabular}
    \caption{$\gamma_1=1.8$, $\gamma_2=2.01$, $z=1$, $L_{\gamma}=5\times10^{51}$erg/sec, $\Gamma=130$, $t_v=25$msec, $\epsilon^b_{\gamma}=0.5$MeV, $\epsilon_B=\epsilon_e=0.3$. Line styles same as in Figure 2}
    \label{test4}
  \end{center}
\end{figure}

\clearpage

\begin{figure}
  \begin{center}
    \begin{tabular}{cc}
      \resizebox{80mm}{!}{\includegraphics{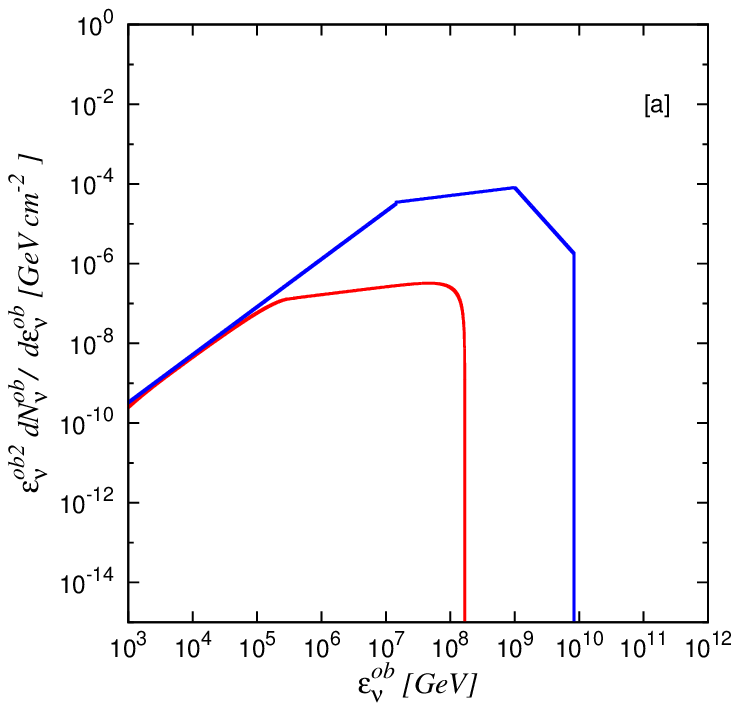}} &
      \resizebox{80mm}{!}{\includegraphics{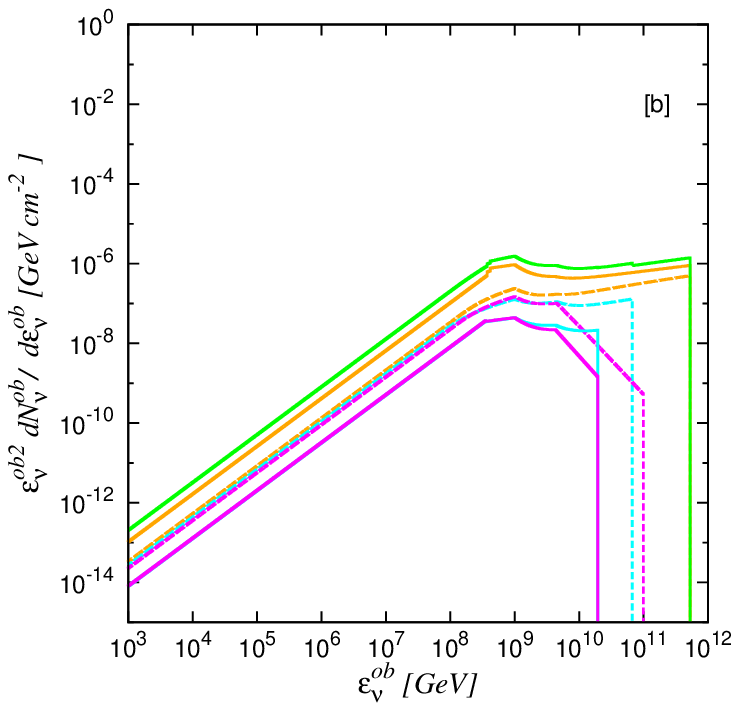}} \\
      \resizebox{80mm}{!}{\includegraphics{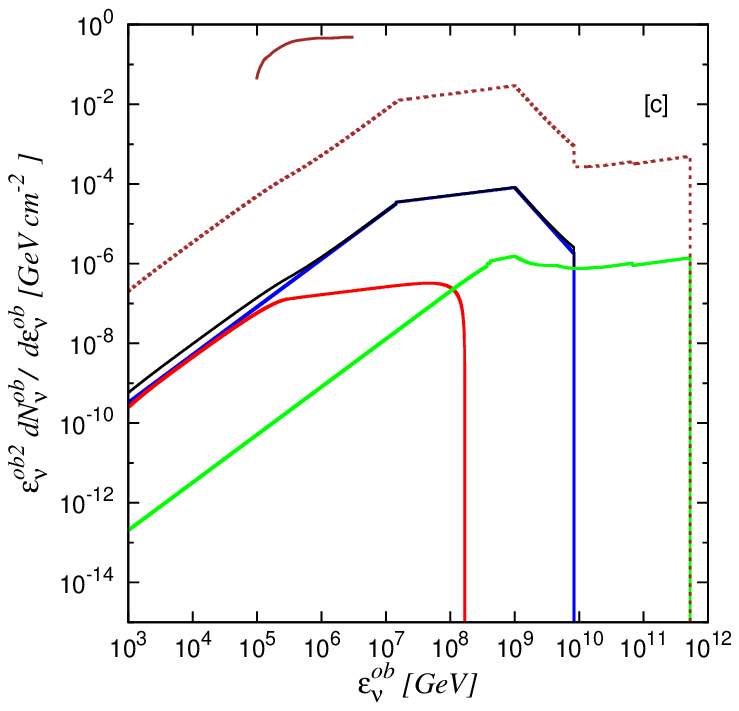}} &
      \resizebox{80mm}{!}{\includegraphics{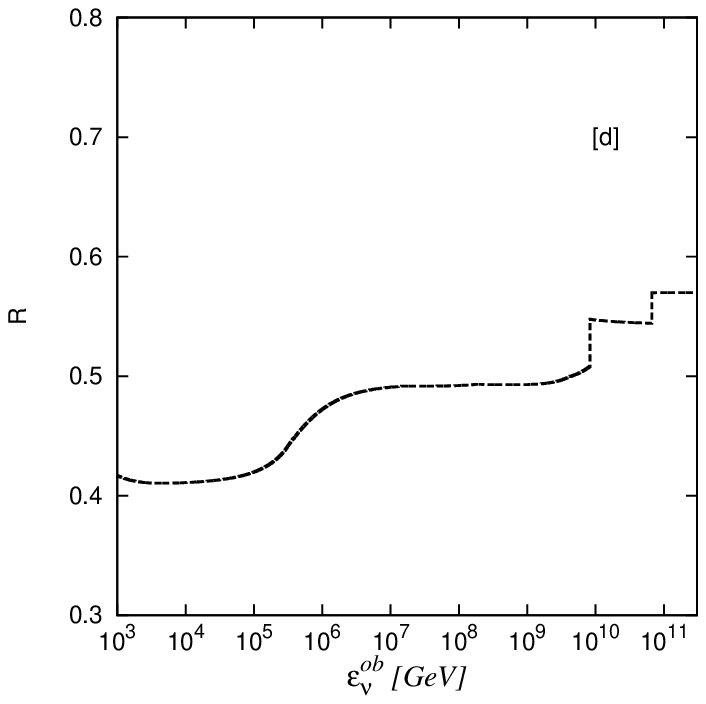}} \\
    \end{tabular}
    \caption{$\gamma_1=1.2$, $\gamma_2=2.2$, $z=0.8$, $L_{\gamma}=10^{52}$erg/sec, $\Gamma=1000$, $t_v=10$msec, $\epsilon^b_{\gamma}=0.5$MeV, $\epsilon_B=\epsilon_e=0.1$. Line styles same as in Figure 2.}
    \label{test4}
  \end{center}
\end{figure}

\clearpage

\begin{figure}
  \begin{center}
    \begin{tabular}{cc}
      \resizebox{80mm}{!}{\includegraphics{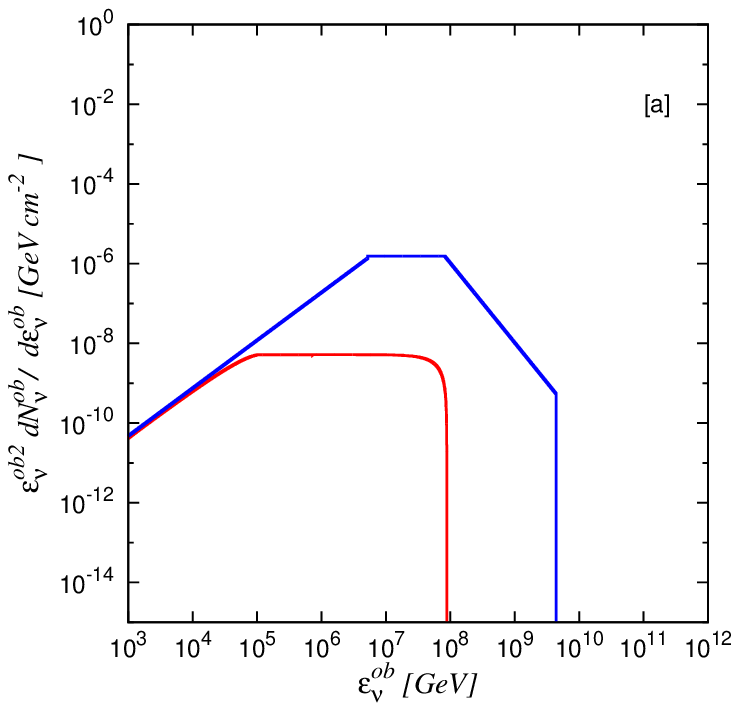}} &
      \resizebox{80mm}{!}{\includegraphics{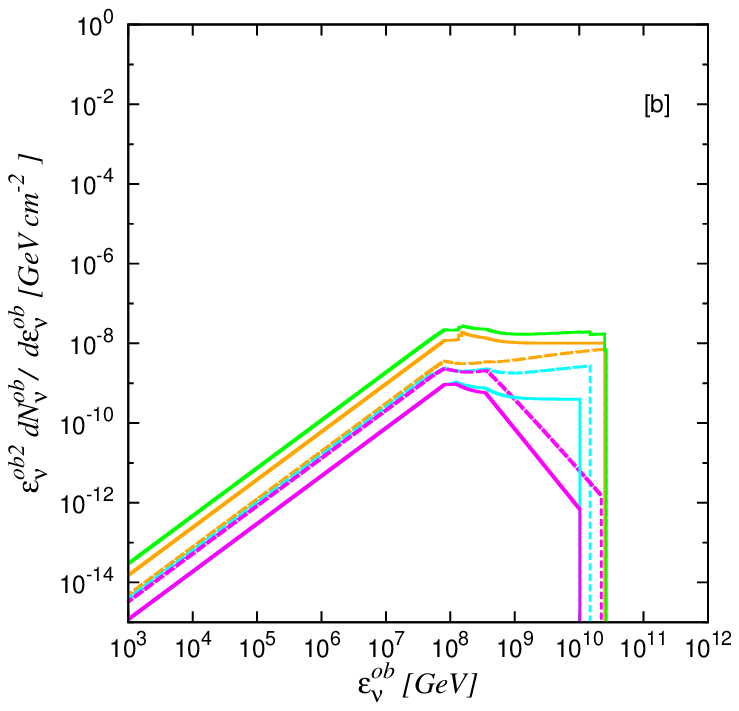}} \\
      \resizebox{80mm}{!}{\includegraphics{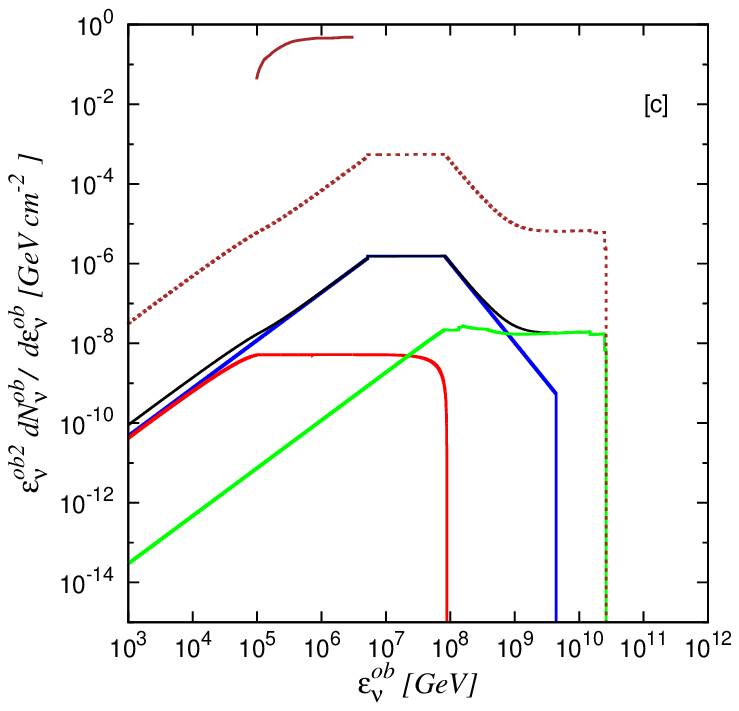}} &
      \resizebox{80mm}{!}{\includegraphics{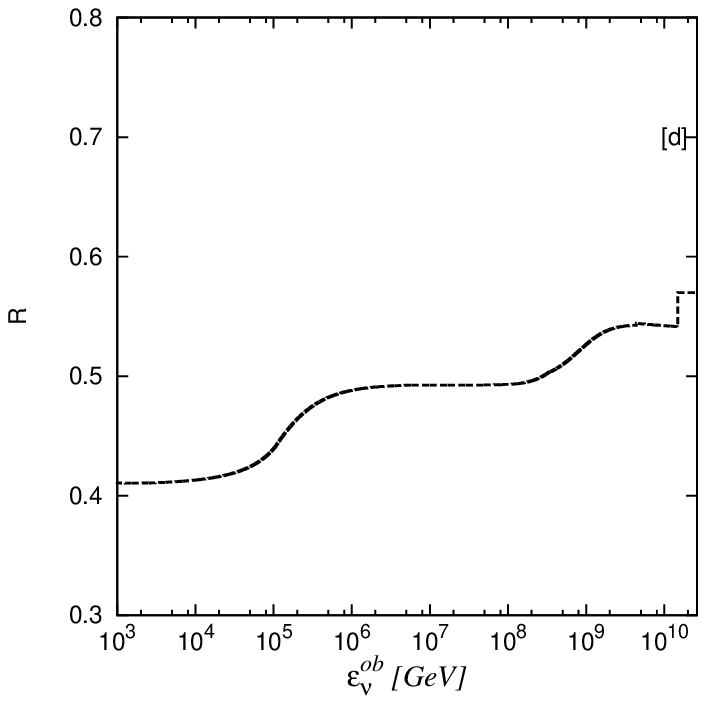}} \\
    \end{tabular}
    \caption{$\gamma_1=1$, $\gamma_2=2.2$, $z=1$, $L_{\gamma}=10^{51}$erg/sec, $\Gamma=600$, $t_v=2$msec, $\epsilon^b_{\gamma}=0.5$MeV, 
$\epsilon_B=\epsilon_e=0.3$. Line styles same as in Figure 2.}
    \label{test4}
  \end{center}
\end{figure}

\end{document}